\documentclass[aps,prl,twocolumn,superscript,superscriptaddress,footinbib,longbibliography]{revtex4-1}
\usepackage{amsmath,amssymb}
\usepackage{bm}
\usepackage{epsfig}
\usepackage{natbib}
\usepackage{hyperref}
\usepackage{color}
\usepackage{graphicx}

\begin{document}
\newcommand{\bsts}{Bi$_{1.5}$Sb$_{0.5}$Te$_{1.7}$Se$_{1.3}$ }
\newcommand{\BS}{Bi$_2$Se$_3$ }
\title{Reducing the impact of bulk doping on transport properties of Bi-based 3D topological insulators}


\author{Shaham Jafarpisheh}\affiliation{2nd Institute of Physics and JARA-FIT, RWTH Aachen University, 52074 Aachen, Germany}\affiliation{Peter Gr\"unberg Institute (PGI-9), Forschungszentrum J\"ulich, 52425 J\"ulich, Germany}
\author{An Ju}\affiliation{2nd Institute of Physics and JARA-FIT, RWTH Aachen University, 52074 Aachen, Germany}
\author{Kevin Jan\ss en}\affiliation{2nd Institute of Physics and JARA-FIT, RWTH Aachen University, 52074 Aachen, Germany}
\author{An Ju}\affiliation{2nd Institute of Physics and JARA-FIT, RWTH Aachen University, 52074 Aachen, Germany}
\author{Takashi Taniguchi}\affiliation{National Institute for Materials Science, 1-1 Namiki, Tsukuba 305-0044, Japan}
\author{Kenji Watanabe}\affiliation{National Institute for Materials Science, 1-1 Namiki, Tsukuba 305-0044, Japan}
\author{Christoph Stampfer}\affiliation{2nd Institute of Physics and JARA-FIT, RWTH Aachen University, 52074 Aachen, Germany}\affiliation{Peter Gr\"unberg Institute (PGI-9), Forschungszentrum J\"ulich, 52425 J\"ulich, Germany}
\author{Bernd Beschoten}\affiliation{2nd Institute of Physics and JARA-FIT, RWTH Aachen University, 52074 Aachen, Germany}
\thanks{ e-mail: bernd.beschoten@physik.rwth-aachen.de}

\begin{abstract}
{
 The observation of helical surface states in Bi-based three-dimentional topological insulators has been a challenge since their theoretical prediction. The main issue raises when the Fermi level shifts deep into the bulk conduction band due to the unintentional doping. This results in a metallic conduction of the bulk which dominates the transport measurements and hinders the probing of the surface states in these experiments. In this study, we investigate various strategies to reduce the residual doping in Bi-based topological insulators. Flakes of \BS and Bi$_{1.5}$Sb$_{0.5}$Te$_{1.7}$Se$_{1.3}$ are grown by physical vapor deposition and their structural and electronic properties are compared to mechanically exfoliated flakes. Using Raman spectroscopy, we explore the role of the substrate in this process and present the optimal conditions for the fabrication of high quality crystals. Despite of this improvement, we show that the vapor phase deposited flakes still suffer from structural disorder which leads to the residual n-type doping of the bulk. Using magneto-measurements we show that exfoliated flakes have better electrical properties and are thus more promising for the probing of surface states.

 }
\end{abstract}

\maketitle   

\section{Introduction}
The relevance of topology in the electronic properties of materials gives rise to the concept of topological insulators (TIs) which is a fascinating frontier in the condensed matter physics \cite{Kosterlitz1973Apr,Kane2005Sep,Kane2005Nov}. The non-trivial band structure of these materials ensures the presence of gapless surface states with a helical spin texture where the spin orientation of the electrons is locked to their momentum direction. Therefore, as long as the time-reversal symmetry is preserved, electrons are protected against backscattering. Such dissipationless current can improve the efficiency of the electronics and are highly desired for future spintronic and photonic devices~\cite{Yan2014Oct,Ivanov2019Dec}. Since the first experimental confirmation of two-dimensional TIs in 2007~\cite{Konig2007Nov}, an extensive effort has been directed toward the observation of surface states in three-dimensional (3D) systems using electronic transport measurements~\cite{Qu2010Aug,Analytis2010Nov,Ren2010Dec,Taskin2011Jun,Kim2011Aug,Sacepe2011Dec,Xiong2012Jul,Taskin2012Aug,Bansal2012Sep,Kim2013Jun,Weyrich2016Oct,Seifert2017Feb}. However, the challenge remains due to the parallel conduction of the bulk states in the 3D crystals and the development of a reliable method to distinguish the contribution of the surface states in the electrical conduction. Unintentional doping of the 3D TIs, specifically in the Bi-based compounds, places the Fermi level deep inside the bulk conduction band. This results in a metallic conduction of the bulk states instead of a semiconducting behavior expected from the gaped band structure. A step forward is to analyze the source of this doping and to improve the fabrication process with the goal of reducing the bulk contribution to the electrical transport.

In this study, we present the vapor phase deposition of Bi-based topological insulators on various substrates including SiO$_2$, hexagonal boron nitride (hBN) and graphene. We present a systematic thickness-dependent Raman spectroscopy study to gain insight into the role of the substrate for  the structural disorder of the fabricated crystals. Next, several approaches to improve the transport properties of the 3D TIs are discussed in detail.

\section{Physical vapor deposition (PVD)}
\begin{figure}[tp]
	\centering
	\includegraphics[width=\linewidth]{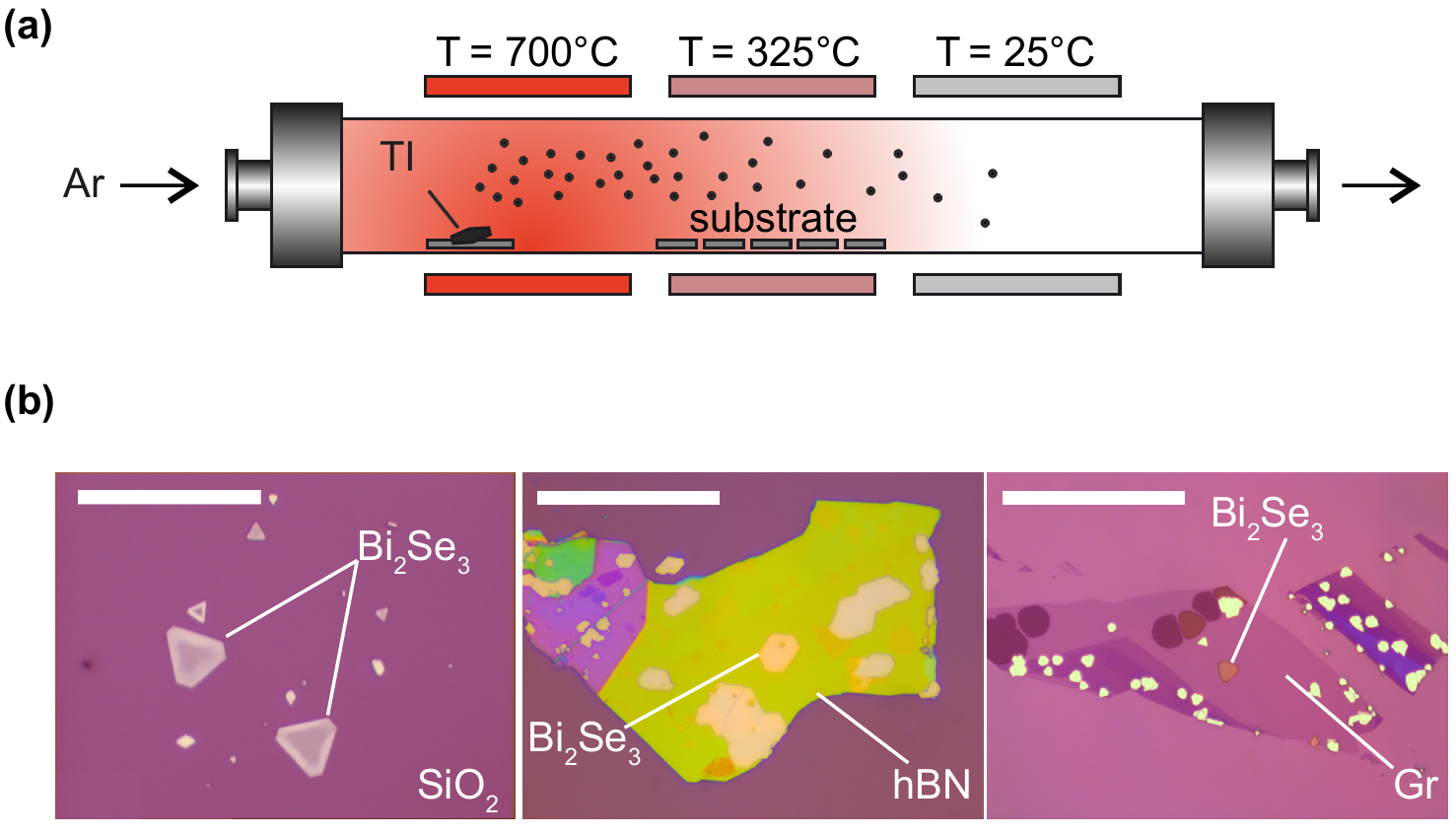}
	\caption{(a) Schematic of the three-zone furnace which is used to deposit thin TI flakes on
		various substrates. (b) Growth of \BS flakes with thicknesses in the range of 1 to 50 nm on SiO$_2$, hBN and graphene (Gr). The scale-bars correspond to a length of 10~$\mu$m.}
	\label{fig:Fig1}
\end{figure}

Vapor-phase growth is a robust technique for the fabrication of TIs on various substrates with the desired thicknesses and forms \cite{Ockelmann2015Aug,Gehring2012Oct,Dang2010Aug,Yan2013Feb,Kong2010Jan}. In this technique, the source material is heated to elevated temperatures and slowly evaporates during the growth process. An inert carrier gas such as Ar carries the evaporated material onto the deposition substrates. The lower temperature in this region results in a highly controllable deposition process. Fig.~\ref{fig:Fig1}a shows a schematic view of the growth chamber which is used in this study. The source crystal is heated to a temperature of T~$\approx$~600-700~$^\circ$C while the substrates are placed at a lower temperature of T~$\approx$~300-350~$^\circ$C.

\begin{figure}[tp]
  	\centering
  	\includegraphics[width=\linewidth]{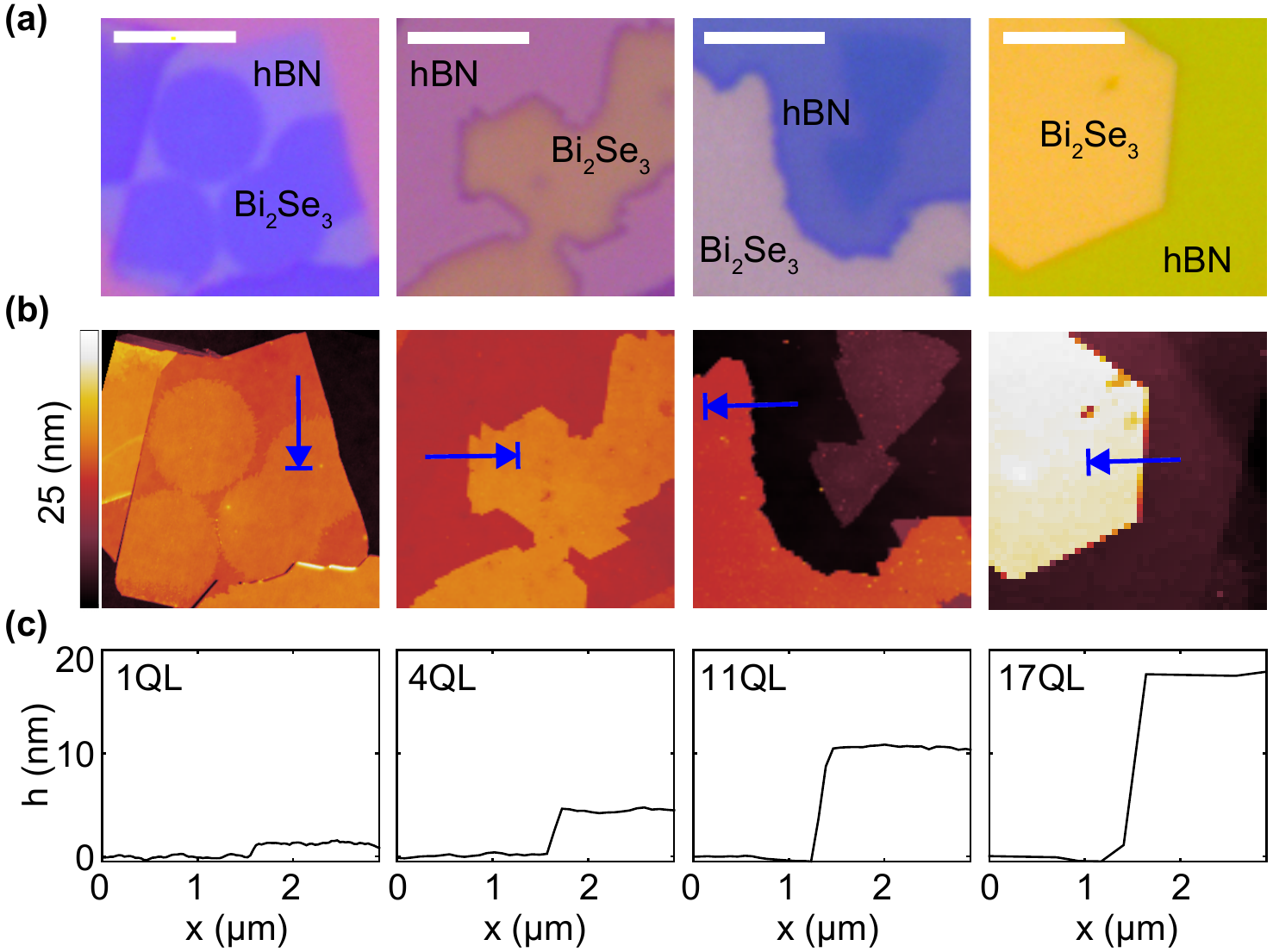}
  	\caption{(a) Optical images of \BS flakes grown with various thicknesses on exfoliated hBN. (b) AFM scan of the flakes shown in part a. (c) AFM line scan at the edge of the flakes as it is depicted in part b showing the thickness (scale bars are 4~$\mu$m).}
  	\label{fig:Fig2}
  \end{figure}

By optimizing the growth process, highly crystalline flakes can be grown on various substrates including SiO$_2$, hBN and graphene (Gr) as it is shown in Fig.~\ref{fig:Fig1}b. Independent of the particular substrate there is a huge spread in the thicknesses of the deposited crystals ranging from 1 to 300 nm in a single run. While this growth mode is favorable when studying thickness dependent structural properties (see next section), it does not allow synthesizing large area flakes with a controlled number of layers. We note that higher control over the layer number can be achieved for ultrathin flakes using solution-based growth methods~\cite{Min2012Feb,Xie2016Aug}. Our PVD technique does not require any catalyst for the deposition which reduces one source of contamination in the process. However, the rough surface of SiO$_2$ can induce structural defects in the TI crystal and increase the doping level of the bulk~\cite{Gehring2012Oct}. One solution is to grow free-standing flakes with minimum contact to the substrate which can reduce the impact of SiO$_2$ surface~\cite{Ockelmann2015Aug}. In this method, however, both surfaces of the TI are exposed to air and water vapor when the flakes are removed from the growth chamber, which can significantly increase unintentional doping. An alternative is the use of hBN as a substrate, which offers an atomically flat surface~\cite{Xue2011Feb}. In this approach, the hBN flakes are exfoliated onto Si/SiO$_2$ substrate prior to the PVD process. Next, TI flakes are grown by PVD directly on hBN (see Fig.~\ref{fig:Fig1}b middle). Thickness dependent color contrasts of optical images, which results from the interference of the reflected light of the flakes, can be used to identify  suitable crystals for transport measurements. This is illustrated in Fig.~\ref{fig:Fig2}a where the color of TI flakes changes from dark blue for ultra-thin crystals to yellow for thicker ones as confirmed by atomic force microscopy (AFM) scans (Figs.~\ref{fig:Fig2}b and \ref{fig:Fig2}c). Therefore, TI flakes with the desired thicknesses can directly be identify after the PVD growth without the need of an additional AFM scan. This significantly reduces the air exposure time prior to both processing and transport measurements.

\section{Thickness-dependent Raman spectroscopy}
\begin{figure*}[tp]
	\centering
	\includegraphics[width=\textwidth]{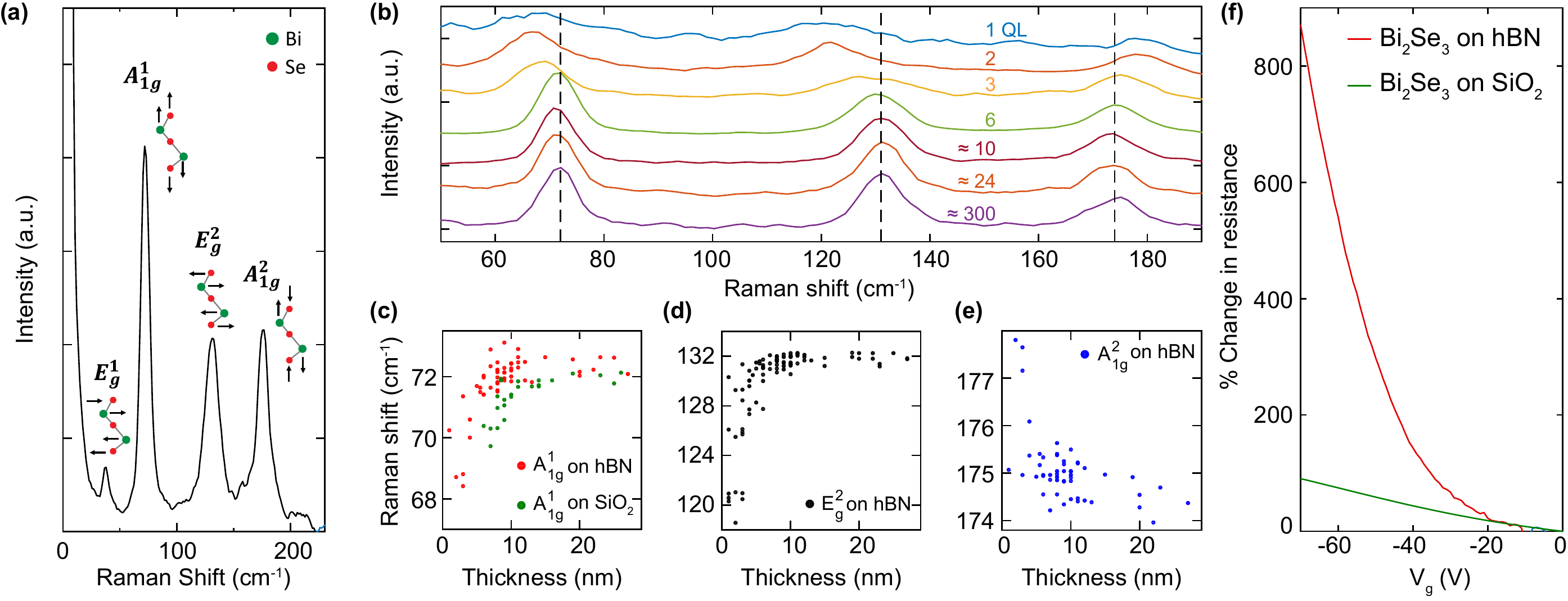}
	\caption{(a) A typical Raman spectrum of \BS and schematic view of each vibrational mode. (b) Thickness dependent Raman spectra of Bi$_2$Se$_3$. (c-d) Peak positions extracted using Lorentzian fits of the Raman signals for (c) A$^1_\mathrm{1g}$ (d) E$^2_\mathrm{g}$ and (f) A$^2_\mathrm{1g}$ vibrational modes. (f) Four-probe resistance as a function of back gate voltage for \BS deposited on hBN and SiO$_2$. Both flakes have similar thicknesses of around $\approx$10~nm.}
	\label{fig:Fig3}
\end{figure*}

\BS has a rhombohedral crystal structure with four Raman active modes, two in-plane (E$_\mathrm{g}$) and two out-of-plane (A$_\mathrm{g}$) modes. Fig.~\ref{fig:Fig3}a shows a typical Raman spectrum of the PVD fabricated flakes with E$^1_\mathrm{g}\sim$ 37 cm$^{-1}$, E$^2_\mathrm{g}\sim$ 131 cm$^{-1}$, A$^1_\mathrm{1g}\sim$ 72  cm$^{-1}$ and A$^2_\mathrm{1g}\sim$ 174 cm$^{-1}$ confirming the composition of the deposited flakes~\cite{Shahil2012Mar,Richter1977Dec,Zhang2011Jun}. The thickness dependency of these Raman modes provides information on the structural properties of the layer and is a tool to identify the role of the substrate~\cite{Shahil2012Mar}. Specifically, we observed that the the strained induced in the TI crystal structure as result of the surface roughness of SiO$_2$ can penetrate several nanometers more than that of hBN.

Fig.~\ref{fig:Fig3}b shows the evolution of the Raman spectra of the \BS flake on hBN with decreasing thicknesses ranging from around 300 to 1 quintuple layer (1~QL $\approx$ 1~nm). The intensity of all spectra are normalized by their A$^2_\mathrm{1g}$ mode and shifted vertically for clarity. By fitting Lorentzian line shapes, the positions of each Raman peak is extracted and plotted as a function of the thickness in Fig.~\ref{fig:Fig3}c-d. The red-shift in A$^1_\mathrm{1g}$ and E$^2_\mathrm{g}$ modes in addition to the blue-shift in A$^2_\mathrm{1g}$ for thicknesses lower than $\approx$10~nm is consistent with previous reports~\cite{Zhang2011Jun,Wang2013Sep,Xu2015Apr}. This shift for thin flakes could be either due to the phonon confinement or due to strain induced from the interface between the flakes and the substrate. Phonon confinement in thin flakes is expected to mainly affect the out-of-plane vibrational modes, while a stronger effect on the in-plane modes is expected from the substrate-induced strain. Based on our observations, the peak position of the in-plane mode (E$^2_\mathrm{g}$ in Fig.~\ref{fig:Fig3}d) is affected more strongly than the respective positions of the out-of-plane modes (A$^1_\mathrm{1g}$ and A$^2_\mathrm{1g}$ in Fig.~\ref{fig:Fig3}c and Fig.~\ref{fig:Fig3}e). This suggests that the strain from the substrate, plays a more important role in the structural changes of thinner flakes. The red-shift of E$^2_\mathrm{g}$ and the blue-shift of A$^2_\mathrm{1g}$ are the result of tensile strain in the lattice structure of Bi$_2$Se$_3$. The strain resulting from the hBN substrate penetrates 5 to 6~nm inside the flakes. However, the penetration depth in flakes deposited on SiO$_2$ is $\approx$3 to 4~nm larger as it can be seen by comparing the red and green dots in Fig.~\ref{fig:Fig3}c. This is a direct consequence of the SiO$_2$ roughness and surface inhomogeneities which lead to less screening from the underlying QLs.

Comparing the electrical conductivity of \BS on the two surfaces further illustrates the crucial role of the substrate for the doping concentration in the TI. Using conventional lithography techniques, Cr/Au contacts were fabricated on two typical flakes deposited on hBN and SiO$_2$ substrates. Since the TI flakes were grown free-standing on the SiO$_2$ substrate, an initial transfer process was necessary to achieve flakes laying flat on the surface. This step is not required for the TIs grown on hBN which results in a cleaner process. Comparing the resistances of the two devices, we could observe a larger back gate tunability of the TI grown on hBN substrate  (compare the percentage change in resistance as shown by the red and green traces in Fig.~\ref{fig:Fig3}f). This is a clear indication that the Fermi level of TI/hBN is closer to the bottom edge of the bulk conduction band and indicates an improvement in the overall transport properties of the TI layer.
\begin{figure}[bp]
	\centering
	\includegraphics[width=\linewidth]{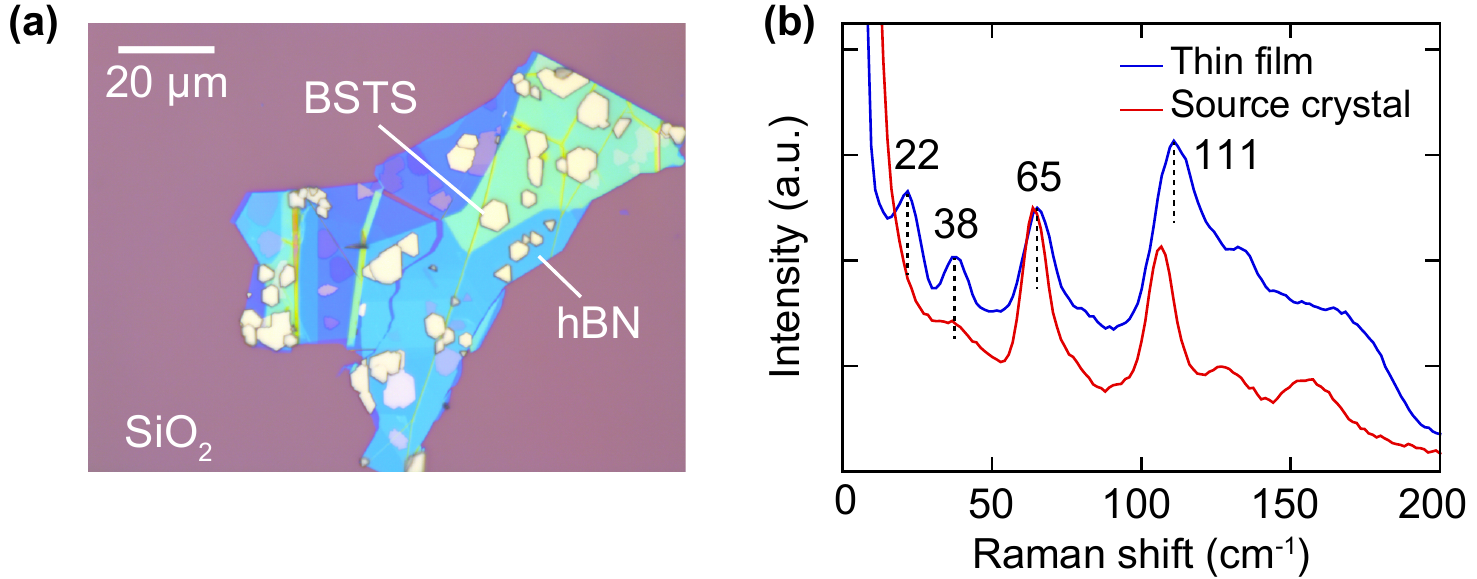}
	\caption{(a) An optical image of BSTS grwon on hBN using the PVD process. (b) Raman spectra of the source crystal used in the PVD process and the deposited thin film.}
	\label{fig:Fig4}
\end{figure}
\begin{figure*}[tp]
	\centering
	\includegraphics[width=\textwidth]{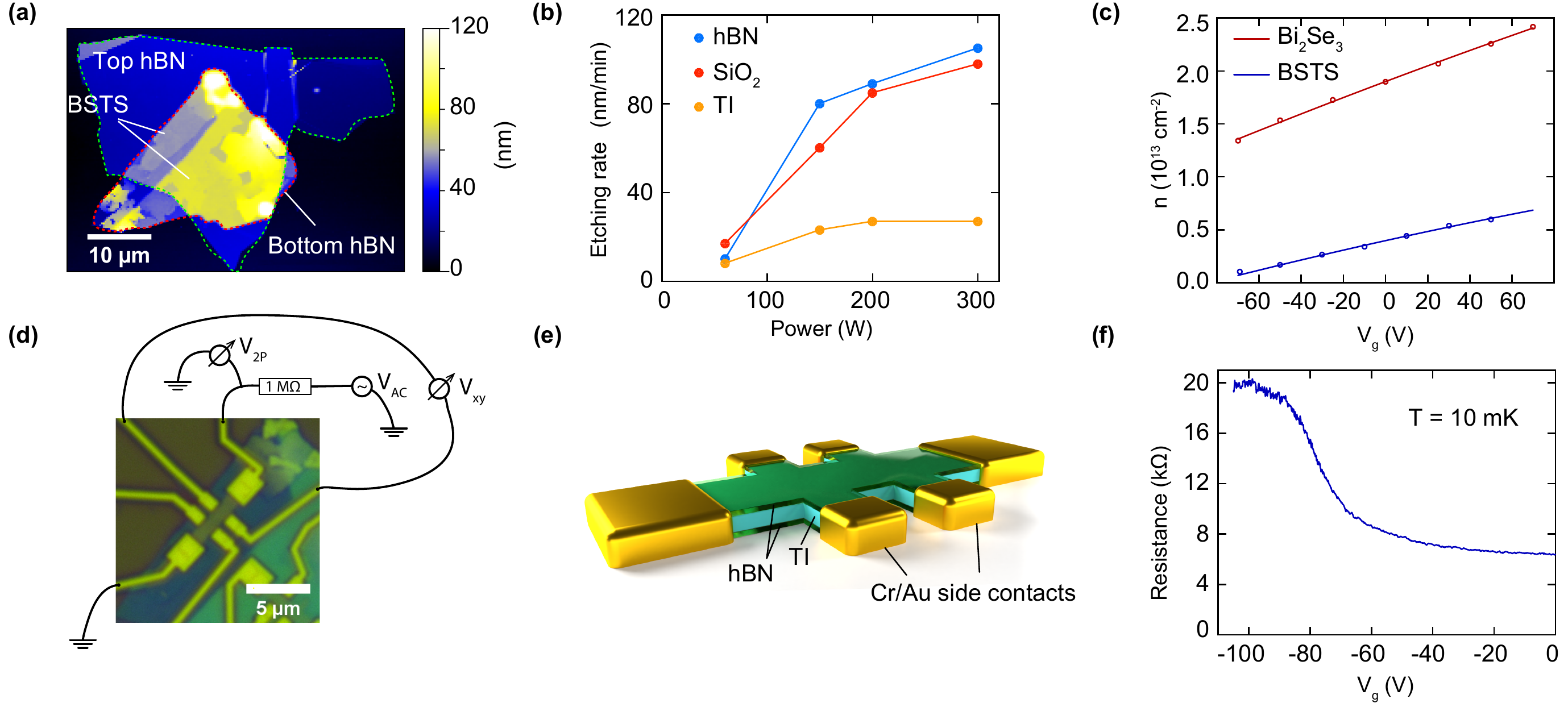}
	\caption{(a) An AFM scan of a PVD grown BSTS encapsulated in hBN. (b) Etching rates of TI, SiO$_2$ and hBN at different RIE powers using CHF$_3$ and Ar gas mixture. (c) Charge carrier density of \BS and BSTS as a function of back gate voltage extracted from Hall measurements. Measurements are performed at $T = 1$~K for \BS and $T = 10$~mK for BSTS. (d) Schematic view of measurement configuration of an encapsulated BSTS Hall bar. (e) View of fully encapsulated TI in hBN and fabricated edge contacts. (f) Four-probe resistance of BSTS as a function of back gate.}
	\label{fig:Fig5}
\end{figure*}

\section{Quaternary TIs}
A promising approach to reduce the large bulk conductivity of binary bismuth-based TIs is the use of quaternary compounds composed of Bi, Sb, Te and Se (BSTS). In these materials negatively charged Se vacancies are compensated by positively charged Te anti-sites when an optimum composition is used. This has been shown to result in a suppression of the bulk residual doping~\cite{Ren2011Oct,Arakane2012Jan}.  Furthermore, ARPES measurements have shown no observable Rashba-splitting in BSTS compounds unlike Bi$_2$Se$_3$~\cite{Arakane2012Jan,Neupane2012Jun,King2011Aug}. This is an advantage of BSTS since it is a challenge to distinguish the contribution of the Rashba-states from the topologically protected states in the transport measurements.

We use the PVD method to deposit thin BSTS flakes on hBN. Fig.~\ref{fig:Fig4}a shows an optical image of such flakes with flat edges which indicate a high crystal quality. Raman spectroscopy can be used to compare the composition of the deposited thin films and the evaporation source material used in this process (see Fig.~\ref{fig:Fig4}b). As confirmed by  energy dispersive X-ray spectroscopy (EDX) measurements (not shown), the source crystals have  composition of Bi$_{1.5}$Sb$_{0.5}$Te$_{1.7}$Se$_{1.3}$.
While the EDX analysis could not be performed on the thin crystals, the similar peak positions of the Raman spectra for the deposited BSTS crystals shown in Fig.~\ref{fig:Fig4}b reveal a comparable composition. The difference between the two spectra is resulted from the thickness-dependent shift of the frequencies similar to our observations in Fig.~\ref{fig:Fig3} for \BS and the measurements reported in Ref.~\cite{Tu2014Aug} for BSTS. Since the air exposure after the deposition can still lead to degradation and doping of the flakes, the top surface of the layers are covered by dry-transferring another hBN flake immediately after the PVD process. This further ensures the protection of the flakes during the fabrication process of electrical contacts. Fig.~\ref{fig:Fig5}a shows an AFM image of a thin BSTS encapsulated in two hBN flakes. In order to electrically access the TI layer, reactive ion etching (RIE) is used to pattern a Hall bar shape through the stack. In this context, it is important to select the etching process gases which allows for high selectivity on hBN and the TI etching while having a minimum effect on the SiO$_2$ layer as strong etching of the  SiO$_2$ layer leads to an unstable back gate during the electrical measurements. After testing various process gases, Ar and CHF$_3$ are found to be the most suitable mixture. As it can be seen in Fig.~\ref{fig:Fig5}b, lower applied RF powers results in similar etching rates of the three layers which is preferential compared to high powers for which SiO$_2$ and hBN are etched at much higher rates compared to the TI layer. The schematic and optical view of the final device are shown in Figs.~\ref{fig:Fig5}d and \ref{fig:Fig5}e. This fabrication process indeed improved the transport properties of the TI layers. Fig.~\ref{fig:Fig5}c shows the charge carrier densities extracted from the Hall effect measurements for both a fully encapsulated BSTS crystal (hBN/BSTS/hBN) (blue data points) and a Bi$_2$Se$_3$/hBN half sandwich (red data points). The charge carrier density of the bulk states is reduced by one order of magnitude (compare blue to red data points). As a result, the backgate tunability of the BSTS flake has significantly improved Fig.~\ref{fig:Fig5}f.

\begin{figure*}[tp]
	\centering
	\includegraphics[width=\textwidth]{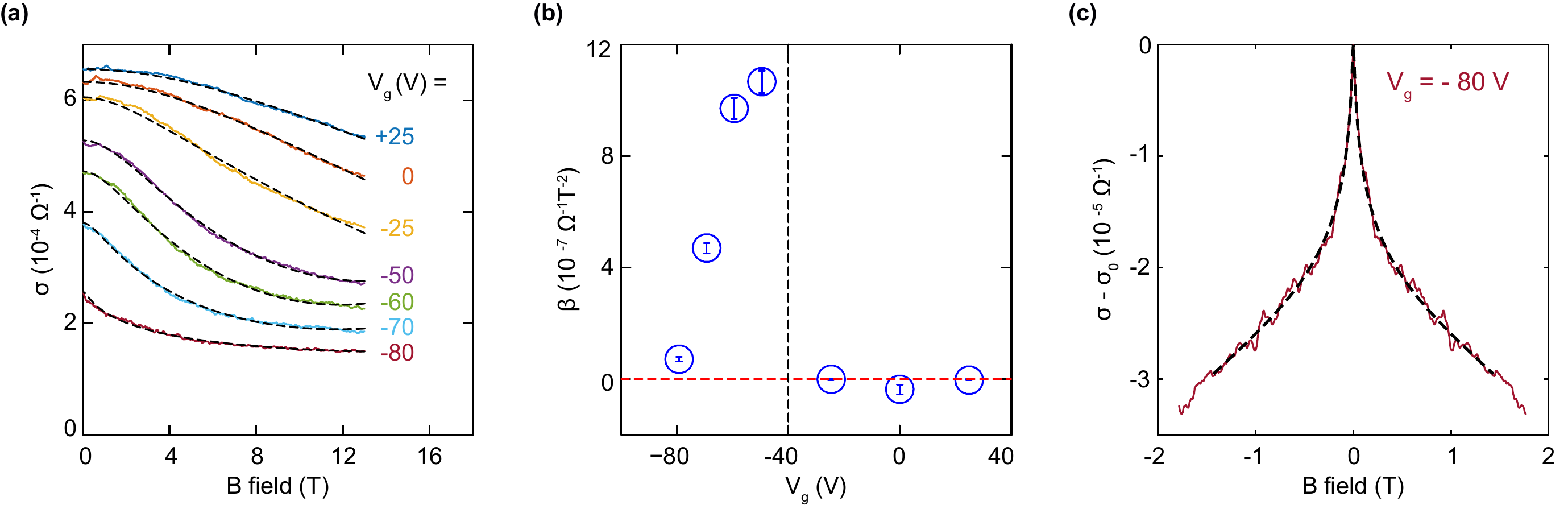}
	\caption{(a) Magneto-conductivity trace of BSTS device measured at different back gate voltages (V$_\mathrm{g}$). The black dashed lines show the modified HLN model. (b) The value of $\beta$ extracted from the HLN fits. The red and black dashed lines indicate $\beta$~=~0 and $V_\mathrm{g}$~=~-40 V, respectively. (c) WAL cusp measured at a back gate voltage of $V_g$ = -80 V. The black dashed line if the fit using HLN model.}
	\label{fig:Fig6}
\end{figure*}

The magneto-conductivity measurements of this device also indicate an improvement of the transport properties. Fig.~\ref{fig:Fig6}a shows the conductivity as a function of perpendicular magnetic field at different back gate voltages. Detailed analysis of the conductivity gives an insight into the role of the bulk states to the transport. The conductivity of the materials with large spin-orbit coupling as a function of magnetic field can be described by the Hikami-Larkin-Nagaoka (HLN) model given as~\cite{Hikami1980Feb}:
\begin{equation}
\label{eq:HLN_modified}
\Delta \sigma = \frac{\alpha e^2}{2 \pi^2 \hbar} \left[ \ln\left( \frac{\hbar}{4 B e l_\phi^2} \right) - \Psi\left(\frac{1}{2}+ \frac{\hbar}{4 B e l_\phi^2} \right)\right] + \beta B^2.
\end{equation}
Here $\alpha$ is a fitting pre-factor, $l_\phi$ the phase coherence length, $\psi$ the digamma function and $B$ is the external magnetic field. The coefficient of the quadratic term ($\beta$), describes additional scattering terms at high magnetic fields and consists of a quantum correction part ($\beta_\mathrm{q}$) and a classical part ($\beta_\mathrm{c}$) as $\beta=\beta_\mathrm{q}+\beta_\mathrm{c}$. The quantum correction term can be described by the following expression~\cite{Hikami1980Feb,Assaf2013Jan}:
\begin{alignat}{2}
\label{eq:QC}
\beta_q&= \frac{- e^2}{48 \pi^2 \hbar} \left[ \frac{1}{B_\mathrm{SO}+B_\mathrm{e}}\right]^2 \nonumber\\
&+ \frac{3e^2}{96 \pi^2 \hbar} \left[ \frac{1}{(4/3)B_\mathrm{SO}+B_{\phi}}\right]^2,
\end{alignat}
with $B_\mathrm{SO,e,\phi} = \hbar/(4el_\mathrm{SO,e,\phi}^2)$. The relative values of both the spin-orbit ($l_\mathrm{SO}$) and elastic scattering ($l_\mathrm{e}$) length scales determine the sign of the quantum correction coefficient. In materials with large spin-orbit coupling such as topological insulators, the $B_\mathrm{SO}$ term ensures the positive sign of the $\beta_\mathrm{q}$ based on the Eq.~(\ref{eq:QC}). On the other hand, the classical part has a negative sign and can be written as~\cite{Assaf2013Jan,Schroder2006}:
\begin{equation}
\label{eq:CC}
\beta_\mathrm{c} = - \mu_\mathrm{GMR}^2\sigma_0.\\
\end{equation}
Here, $\mu_\mathrm{GMR}$ is the geometrical magnetoresistance mobility, $\mu_\mathrm{GMR}=\xi\mu_\mathrm{H}$ with $\mu_\mathrm{H}$ being the Hall mobility and $\xi$ the magnetoresistance scattering factor~\cite{Schroder2006}. Therefore, the sign of the quadratic term in Eq.~(1) indicate whether the classical term and therefore bulk states dominates over the transport. Fig.~\ref{fig:Fig6}b illustrates the values of $\beta$ extracted from the fits shown by the dashed lines in Fig.~\ref{fig:Fig6}a at different back gate voltages. The negative value at zero back gate voltage shows the dominance of bulk transport contributions. However, the sign of $\beta$ becomes positive for negative back gate voltages which is consistent to the lowering of the Fermi energy and a suppression of the bulk conduction channel. This can also qualitatively be seen in Fig.~\ref{fig:Fig6}a by comparing the curvature of the magneto-conductivity traces at high magnetic fields at $V_\mathrm{g}$~=~-25~V (yellow curve) and $V_\mathrm{g}$~=~-50~V (purple curve). Further increase of the back gate voltage results in a smaller values of $\beta$. This is due to the development of the weak antilocalization (WAL) signal which becomes dominant at very low magnetic fields (Fig.~\ref{fig:Fig6}c) and as a result, the first term of Eq.~(\ref{eq:HLN_modified}) gain importance in the fitting procedure. The result of this fit is shown by the dashed line in Fig.~\ref{fig:Fig6}c. Nevertheless, the curvature of magneto-conductivity at $V_\mathrm{g}$~=~-80~V is completely different from $V_\mathrm{g}$~=~+25~V which shows that in these materials it is possible to reduce the contribution of the bulk states using electric gate fields. This is in contrast to \BS where the linear magneto-conductivity is typically observed at all gate voltages (see Fig.~\ref{fig:Fig7}a).
Despite these overall improvements, the PVD grown quaternary TIs still exhibit residual bulk doping. This might also result from large-scale structural defects which occur during the deposition process as observed in epitaxial TI films~\cite{Hickey2018Aug}.

\section{Exfoliation}
Although the PVD method could be important for the fabrication of larger scale TI crystals, it still has major limitations. In comparison with the solution-based methods, PVD offers a lower control over the number of the layers~\cite{Xie2016Aug}. Furthermore, despite major improvements of the transport properties by using quaternary encapsulated PVD layers, exfoliated flakes from single crystals still exhibit the highest electronic quality.

\begin{figure}[tp]
	\centering
	\includegraphics[width=\linewidth]{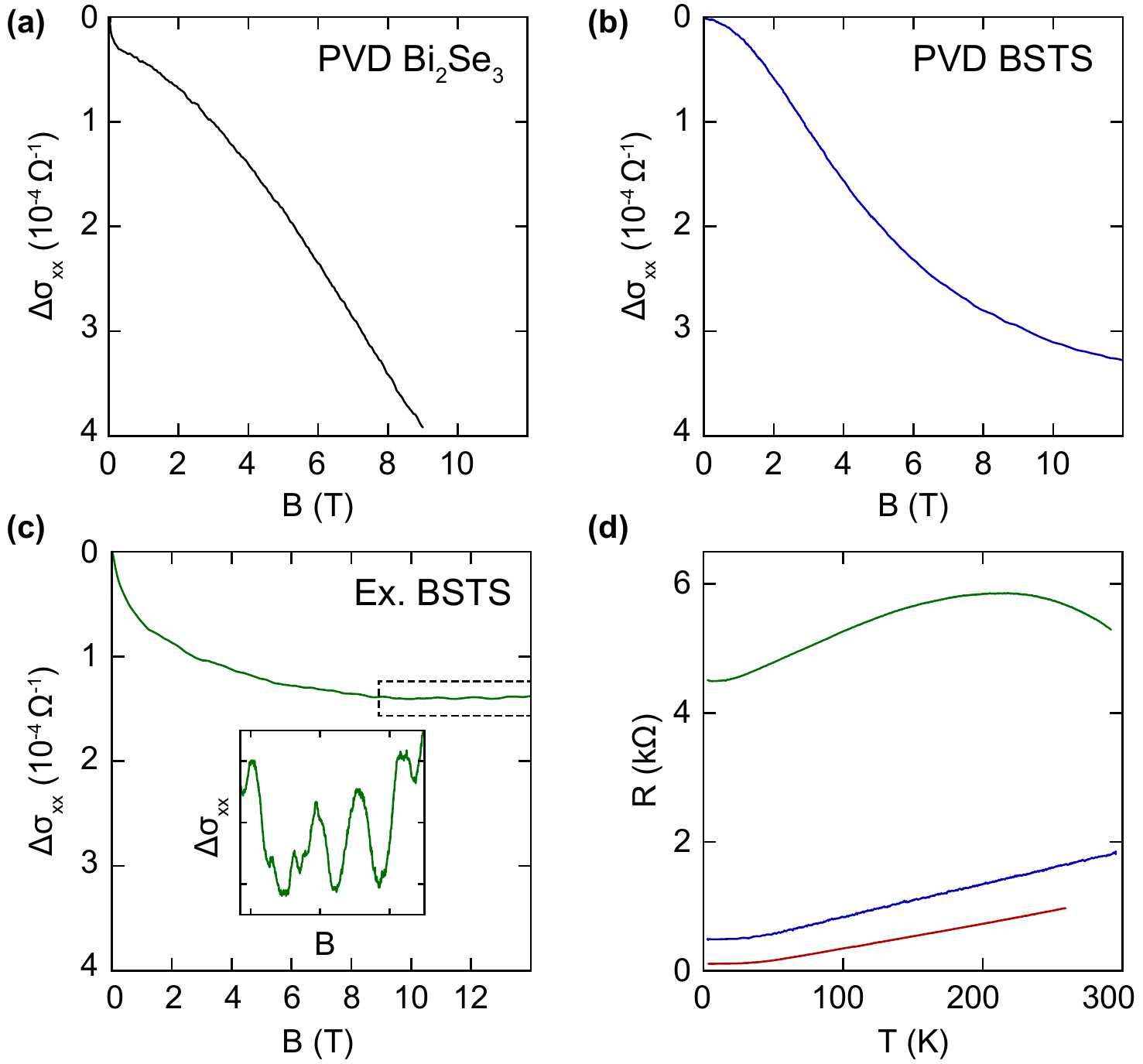}
	\caption{Magneto-conductivity of (a) PVD \BS (b) PVD BSTS and (c) exfoliated BSTS flakes. (d) Temperature dependency of the resistance for the material systems shown in panels (a)-(c).}
	\label{fig:Fig7}
\end{figure}

A direct comparison of the transport properties of \BS and BSTS flakes fabricated by PVD and an exfoliated BSTS flake is shown in Fig.~\ref{fig:Fig7}. As mentioned above, the PVD fabricated \BS (Fig.~\ref{fig:Fig7}a) typically shows a classical linear magneto-conductivity at large magnetic fields due to the large bulk charge carrier densities~\cite{Ockelmann2015Aug,Hu2008Sep}. The dominance of bulk transport can also be observed in the temperature dependent resistance of PVD \BS which shows a metallic behavior in the entire range of the measurements (see the red trace in Fig.~\ref{fig:Fig7}d). Using BSTS as the source material in the PVD process improves the transport properties as discussed in the previous section. The temperature dependency of the resistance in these types of devices is still metallic but with higher resistances. However, the magneto-resistance at high magnetic fields shows quadratic dependency as seen in Fig.~\ref{fig:Fig7}b, which suggests the decoupling of bulk from the surface states~\cite{Singh2017Nov}. 
On the other hand, a major improvement of the transport properties is observed when using the exfoliation technique. The absence of structural defects resulting from the PVD process leads to a lower bulk charge carrier concentration, a larger Hall mobility reaching 4,500 cm$^2$/(Vs) (compared to typical values lower than 1,000 cm$^2$/(Vs) for PVD-grown crystals) and the observation of the quantum oscillations at high magnetic fields (inset of Fig.~\ref{fig:Fig7}c). Furthermore, the semiconducting temperature dependency of the resistance at temperatures close to the room temperature (see the yellow trace in Fig.~\ref{fig:Fig7}d at $T>200$~K) also indicate a smaller bulk contribution in the transport.

We conclude that exfoliated BSTS TIs provide the most suitable material to explore the transport properties of the topologically protected surface states. However, even for this material it is not easily possible to separate the remaining contribution of the bulk states in the measurements such as WAL or the Shubnikov de Haas (SdH) oscillations. In-fact, extracting a $\pi$-Berry phase from the analysis of high $B$-field quantum oscillations with the goal of indicating surface transport can be misleading~\cite{Ren2010Dec,Qu2010Aug,Taskin2011Jun,Analytis2010Nov,Sacepe2011Dec,Xiong2012Jul,Taskin2012Aug}. Based on the study performed by Kuntsevich and co-workers~\cite{Kuntsevich2018May}, other mechanisms such as sample inhomogeneities and a magnetic field dependence of the chemical potential can result in a phase shift similar to what is expected from the surface states. On the other hand, the uncertainty in such analysis is much larger than the extracted values for the phase of the SdH oscillations. Alternatively, spin-sensitive potentiometric measurements in which ferromagnetic electrodes are used to directly detect the spin polarization of the surface current~\cite{Schwab2011Mar,Li2014Feb}. Performing this type of measurements in a non-local detection scheme makes it possible to separate the contribution of the bulk and solely characterize the topologically protected surface states~\cite{Jafarpisheh2019Jul}.

\section{Conclusion}In summary, the fabrication of Bi-based TIs using PVD and exfoliation techniques is discussed in detail. The quality of the crystals and the impact of the substrates has been analyzed using Raman spectroscopy  measurements. Exfoliation from bulk crystals has been identified as a more suitable approach for studying the surface states of TIs compared to samples deposited by the PVD process.

\paragraph*{Acknowledgments.}
This work was supported by the Deutsche Forschungsgemeinschaft (DFG, German Research Foundation) via SPP 1666 (BE 2441/8), and under Germany’s Excellence Strategy - Cluster of Excellence Matter and Light for Quantum Computing (ML4Q) EXC 2004/1 - 390534769, the European Union’s Horizon 2020 research and innovation programme under grant agreement No 785219 (Graphene Flagship), the Virtual Institute for Topological Insulators (J\"ulich- Aachen-W\"urzburg-Shanghai) and by the Helmholtz Nano Facility~\cite{Albrecht2017} at the Forschungszentrum J\"ulich. Growth of hexagonal boron nitride crystals was supported by the Elemental Strategy Initiative conducted by the MEXT, Japan, A3 Foresight by JSPS and the CREST (GrantNo. JPMJCR15F3), JST.

\bibliographystyle{pss}
\providecommand{\WileyBibTextsc}{}
\let\textsc\WileyBibTextsc
\providecommand{\othercit}{}
\providecommand{\jr}[1]{#1}
\providecommand{\etal}{~et~al.}

\end{document}